\documentclass[showpacs,aps,pre,floatfix,superscriptaddress]{revtex4}
\usepackage{bm} 
\usepackage{amssymb}
\usepackage{natbib}
\usepackage{graphicx}
\usepackage{color}
\usepackage{citesort}

\usepackage{amsmath}

\newsavebox{\astrutbox}
\sbox{\astrutbox}{\rule[-5pt]{0pt}{20pt}}

\newcommand{\del}{\nabla}

\newcommand{\be}{\begin{equation}}
\newcommand{\ee}{\end{equation}}
\newcommand{\bea}{\begin{eqnarray}}
\newcommand{\eea}{\end{eqnarray}}
\newcommand{\pat}{\partial}

\newcommand\btau{\boldsymbol{\tau}}

\renewcommand\div{\boldsymbol{\nabla}}

\newcommand{\dcn}{{\rm d}{\cal C}^N}
\newcommand{\dcnm}{{\rm d}{\cal C}^{N-1}}

\begin{document}

\title{Effective velocity boundary condition at a 
mixed slip surface}

\author{M. Sbragaglia} \affiliation{Department of Applied Physics, University of Twente, P.O. Box 217, 7500 AE, Enschede, The Netherlands.}
\author{A.Prosperetti} \affiliation{Department of Applied Physics, University of Twente, P.O. Box 217, 7500 AE, Enschede, The Netherlands.} \affiliation{Department of Mechanical Engineering, The Johns Hopkins 
University, Baltimore MD 21218, USA}
\date{\today}

\begin{abstract}
This paper studies the nature of the effective velocity boundary conditions 
for liquid flow over a plane boundary on which small free-slip islands 
are randomly distributed. 
It is found that, to lowest order  in the area fraction $\beta$ covered by 
free-slip regions with characteristic size $a$, a macroscopic Navier-type 
slip condition emerges with a slip length of the order of $a\beta$. 
The study is motivated 
by recent experiments which suggest that gas nano-bubbles may form on 
solid walls and may be responsible for the appearance of a partial slip 
boundary conditions for liquid flow. The results are also relevant for 
ultra-hydrophobic surfaces exploiting the so-called ``lotus effect''. 
\end{abstract}


\maketitle

\section{Introduction}

The recent blossoming of research in micro-fluidics has prompted a renewed 
interest in the possibility of slip boundary conditions at the contact of 
a liquid with a solid wall  \cite{LaugaStone03,Laugaetal05}.
While many experiments have provided evidence for a violation of the 
classical no-slip boundary condition at small spatial scales  \cite{Vinogradova99,Watanabeetal99,Pitetal00,ChengGiordano02,Craigetal01,ZhuGranick01a,ZhuGranick02,TrethewayMeinhart02}
the physical mechanisms responsible for this phenomenon are still 
unclear. An interesting possibility is the recent discovery of what appear 
to be small gas nano-bubbles or pockets attached to the wall 
\cite{Bunkinetal96,Ishidaetal00,TyrrellAttard01,Holmbergetal03,Steitzetal03,Simonsenetal04,DammerLohse06}.
The evidence for the existence of these nano-bubbles is somewhat indirect, 
but nevertheless compelling. 
It is also hypothesized and, sometimes, experimentally verified (see Watanabe 
et al. \cite{Watanabeetal99}), that gas pockets may form in cracks or other imperfections of the solid wall, thereby decreasing the overall wall stress. 

In order to explore the macroscopic consequences of the existence of such 
drag-reducing gaseous structures on a solid wall, in this study we consider 
by statistical means the effective velocity boundary condition produced by 
a random distribution of small free-slip regions on an otherwise no-slip 
boundary. 
We consider both the three-dimensional problem, in which the 
regions are equal disks, and the two-dimensional problem, in 
which they are strips  oriented perpendicularly to the flow. 
While idealized, these geometries provide some insight into the macroscopic 
effects of randomly distributed microscopic free-slip regions. 

We find that, away from the wall, the velocity field appears to satisfy a 
partial-slip condition with a slip length proportional, to leading 
order, to the product of the length scale $a$ of the free-slip islands 
and the area fraction $\beta$ covered by them. After 
deriving a general result, we solve the problem to first order accuracy 
in $\beta$ for both the two- and three- dimensional situations. 

As discussed in section \ref{concl}, our results are consistent with those 
of a recent paper by 
Lauga \& Stone \cite{LaugaStone03}, who assumed a periodic distribution of free-slip 
patches on a boundary, as well as those of an older paper by Philip \cite{Philip72}  who 
similarly investigated the effect of free-slip strips arranged periodically 
on a plane wall parallel or orthogonal to the direction of the flow.

The present results are also related to so-called ``lotus effect'' 
\cite{BarthlottNeinhaus97} exploited to obtain ultra-hydrophobic surfaces. 
Such surfaces are manufactured by covering a solid boundary with an array of 
hydrophobic micron-size posts which, due to the effect of surface tension, 
prevent a complete wetting of the wall  
\cite{Ouetal04,OuRothstein05,ChoiKim06}.
In the space between the posts the 
liquid remains suspended away from the wall with its surface in 
contact only with the ambient gas and a concomitant reduction in 
the mean traction per unit area. 
Another instance of drag reduction by a similar mechanism has also been reported  in Watanabe et al. \cite{Watanabeetal99}. These authors studied the pressure 
drop in the flow of a viscous liquid in a tube the wall of which contained 
many fine grooves which prevented a complete wetting of the boundary. 

The approach used in this paper is mainly suggested by the theory of multiple 
scattering \cite{Foldy45,Twersky57,Twersky83} 
and was used before to derive the effective  boundary 
conditions at a rough surface for the Laplace and Stokes problems  \cite{SarkarProsperetti95,SarkarProsperetti96}

\section{Formulation}
\label{formu}

We consider the flow in the neighborhood of a locally plane boundary \footnote{
For the present purposes a curved boundary can be considered plane provided 
the radius of curvature is large compared with the size of the free-slip 
regions and their mean reciprocal distance, of order $a/\beta^{1/2}$.}  
${\cal B}$ with a composite micro-structure which dictates free-slip 
conditions on certain areas $s^1,\,s^2,\ldots,\,s^N$ and no-slip conditions 
on the remainder ${\cal B}-\cup_{\alpha=1}^Ns^\alpha$ (figure \ref{sketch}). 
If each ``island'' 
$s^\alpha$ is sufficiently small, and $\cup_{\alpha=1}^Ns^\alpha$ is also 
sufficiently small (both in a sense to be made precise later), 
near the boundary the flow is described by the Stokes equations: 
\be
 \div p \,=\, \mu \del^2 {\bm u}\, , \qquad \div\cdot{\bm u}\,=\, 0 \, ,
\label{stokeq}
\ee 
in which $p$ and ${\bm u}$ are the pressure and velocity fields and $\mu$ the 
viscosity. On the free-slip regions ${\bm u}$ satisfies the condition of 
vanishing tangential stress:
\be
 {\bm t}_J \cdot ( \btau \cdot \hat {\bm n} ) \,=\, 0 \qquad  {\bm x}\in s^\alpha \qquad 
\alpha\,=\, 1,\,2,\, \ldots,\,N \, \qquad J=2,\, 3 
\label{slip}
\ee
where ${\bm t}_2$ and ${\bm t}_3$ are two unit vectors in the plane and 
${\bm \tau}$ the viscous stress tensor, while, on the rest of the surface,  
\be
  {\bm u}\,=\, 0 \qquad {\bm x}\not\in \cup_{\alpha=1}^N s^\alpha \, .
\ee
The normal velocity vanishes everywhere on ${\cal B}$.

\unitlength 1cm
\begin{figure}
\begin{picture}(15,7)(1.5,0) 
\put(1.8,1){\includegraphics{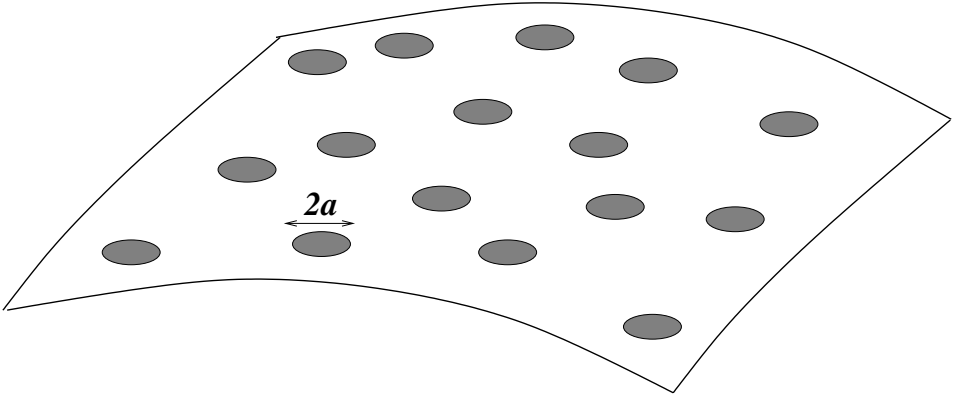}}
\end{picture}
\caption{Solid no-slip boundary with a random distribution of equal circular 
free-slip areas. }
\label{sketch}
\end{figure}

We start by decomposing the solution $(p,\,{\bm u})$ as 
\be
 {\bm u}\,=\,{\bm u}^0+\sum_{\alpha=1}^N{\bm v}^\alpha \, , \qquad
p\,=\,p^0+\sum_{\alpha=1}^N q^\alpha
\label{deco}
\ee
Here ${\bm u}^0$ and $p^0$ are the (deterministic) solution satisfying the 
usual no slip condition on the entire boundary ${\cal B}$ while the fields 
$(q^{\alpha},{\bm v}^{\alpha})$ account for the effect of 
the $\alpha$-th island. We define these local fields so that 
${\bm v}^\alpha$ vanishes everywhere on 
${\cal B}$ except on $s^\alpha$, where it is such that the free-slip 
condition (\ref{slip}) is satisfied. To express this condition it is 
convenient to define 
\be
 {\bm w}^\alpha \,=\, {\bm u}^0 +\sum_{\beta \ne\alpha}{\bm v}^\beta \, , 
\qquad  r^\alpha \,=\, p^0 +\sum_{\beta \ne\alpha}q^\beta \, , 
\ee
so that, for every $\alpha$ = 1, 2, $\ldots$, $N$,
\be
 {\bm u}\,=\, {\bm v}^\alpha +{\bm w}^\alpha \,, \qquad 
p\,=\,q^\alpha+r^\alpha \, .
\ee
On $s^\alpha$, then, ${\bm v}^\alpha$ satisfies
\be
 {\bm t}_{J}\cdot ({\bm \tau}^{v\alpha} \cdot \hat{\bm n}) \, = -\,  
{\bm t}_{J} \cdot ({\bm \tau}^{w\alpha} \cdot \hat {\bm n})
\qquad {\bm x}\in s^\alpha \hspace{.2in} J=2,3
\ee
where
\be
{\bm \tau}^{v\alpha}\,=\, \mu\left[ \div {\bm v}^\alpha + 
(\div {\bm v}^\alpha)^T \right] \, , \qquad 
  \tau^{w\alpha}\,=\, \mu\left[ \div {\bm w}^\alpha + 
(\div {\bm w}^\alpha)^T \right] \, ,
\ee
the superscript $T$ denoting the transpose. Clearly
\be
  {\bm v}^\alpha \rightarrow 0 \, , \qquad q^\alpha \rightarrow 0 \qquad 
{\rm as} \qquad |\bm x -\bm y^\alpha| \rightarrow \infty \, 
\ee
with ${\bm y}^{\alpha}$ a  reference point on the $\alpha$-th island. 
It is evident that both  fields ${\bm v}^\alpha$  and ${\bm w}^\alpha$ 
satisfy the Stokes equations. 
In the terminology of multiple scattering, they are often referred to as the 
'scattered' and 'incident' fields, respectively \cite{Foldy45,RubinsteinKeller89}.

\section{Averaging}

We assume that the free-slip islands are identical circular disks with radius 
$a$, centered at ${\bm y}^\alpha$, with $\alpha$ = 1, 2, $\ldots,\, N$. 
We make use of the method of ensemble averaging and consider an ensemble  
of surfaces differring from each other only in the arrangement of the 
$N$ free-slip islands. Each arrangement, or configuration, is denoted by 
${\cal C}^{N}=({\bm y}^{1},{\bm y}^{2},...,{\bm y}^{N})$. A particular 
configuration will then occur with a probability $P({\cal C}^{N})=P(N)$ 
normalized according to: 
\be
  {1\over N!} \int d^2 y^1 \ldots  \int d^2 y^N 
P({\bm y}^1, \ldots, {\bm y}^N) \, \equiv \, 
  {1\over N!} \int d{\cal C}^N P(N) \,=\, 1 \, .
\ee

The ensemble-average velocity is defined as 
\be
\langle{\bm u}\rangle({\bm x})\,=\,{1\over N!}\int \dcn P(N)\,
{\bm u}({\bm x}|N) 
\label{defav}
\ee
where the notation ${\bm u}({\bm x}|N)$ stresses the dependence of the exact 
field not only on the point ${\bm x}$, but also on the configuration of the 
$N$ islands. In view of the fact that 
${\bm u}^0$ is independent of the configuration of the 
disks, substitution of the decomposition (\ref{deco}) into (\ref{defav}) 
gives 
\be
\langle{\bm u}\rangle({\bm x})\,=\,{\bm u}^0({\bm x})+ 
{1\over N!} \sum_{\alpha=1}^N \int \dcn P(N)\, {\bm v}^\alpha({\bm x}|N) \, .
\ee
Since the disks are identical, each one gives the same contribution to 
the integral. Upon introducing the conditional probability 
$P(N-1|{\bm y}^1)$ defined so that $P(N)\,=\,P({\bm y}^1)\,P(N-1|{\bm y}^1)$, 
we may therefore write 
\bea
\langle{\bm u}\rangle({\bm x})\,
&=&\,{\bm u}^0({\bm x})+ 
{1\over (N-1)!}\int \dcn P({\bm y}^1)\,P(N-1|{\bm y}^1)\,  {\bm v}^1({\bm x}|
{\bm y}^1,N-1)
\eea
or, in terms of the conditional average 
\be
 \langle{\bm v}^1\rangle_1({\bm x}|{\bm y}^1)\,=\,{1\over (N-1)!}\int
\dcnm P(N-1|{\bm y}^1)\,  {\bm v}^1({\bm x}|{\bm y}^1,N-1) \, ,
\ee
\be
\langle{\bm u}\rangle({\bm x})\,= \,{\bm u}^0({\bm x})+ 
\int_{\cal B}  d^2 y\, P({\bm y})\,  \langle{\bm v}\rangle_1({\bm x}|
{\bm y}) 
\label{exp39}
\ee
where the integral is over the entire boundary. For convenience, here and in 
the following, we drop the superscript $1$ on the quantities referring to 
disc $1$. 
Since ${\bm v}^\alpha$ and $q^\alpha$  satisfy the Stokes equations 
everywhere, so do $\langle{\bm v}\rangle_1$ and  
$\langle q\rangle_1$. The boundary conditions are 
\be
  \langle{\bm v}\rangle_1 \,=\, 0 \qquad {\bm x}\notin s
\ee 
while
\be
 {\bm t}_J\cdot (\langle {\bm \tau}^v \rangle_{1} \cdot \hat{\bm n}) \, = -\,  
{\bm t}_{J} \cdot (\langle {\bm \tau}^w \rangle_{1} \cdot \hat{\bm n})
\qquad {\bm x}\in s \hspace{.2in} J=2,3 \, . 
\ee
Note that  
\be
 \langle\tau^w_{jk}\rangle_1 \,=\, \mu \left(  \langle\pat_j  w_k 
\rangle_1 +  \langle \pat_k  w_j \rangle_1\right) \, =\, 
\mu \left( \pat_j   \langle w_k 
\rangle_1 + \pat_k  \langle  w_j \rangle_1\right)\, ,
\ee
and similarly for $ \langle\tau^v_{jk}\rangle_1$ 
since averaging and differentiation commute as is evident from the 
definition (\ref{defav}). The normal velocity vanishes everywhere: 
\be
  \langle v_\perp\rangle_1 \,\equiv\, \hat {\bm n}\cdot \langle{\bm v}\rangle_1 
\,=\, 0 \qquad {\bm x}\in {\cal B} \, .
\ee

It may be noted that $ P({\bm x})$ is just the number density of free-slip 
islands per unit surface area of the boundary; the area fraction $\beta$ 
covered by these islands is 
\be
  \beta(\bm x)\,=\,\int_{|\bm x-\bm y|\leq a}P(\bm y) \,d^2y \, \simeq \, 
 \pi a^2P(\bm x) + O(a^2/L^2)
\ee
where $L$, assumed much greater than $a$, is the characteristic length scale 
for variations of the number density.  

The framework just described can be readily extended to disks of unequal
radius, and to non-isotropic islands such as ellipses. In both cases 
the probability density would depend on a suitably enlarged list of 
variables such as the disk radius, the characteristic size, orientation and 
aspect ratio of the ellipses, and so on.

\section{The effective boundary condition}
\label{effbcs}

Now we derive a formal expression for the effective boundary condition on 
${\cal B}$. To this end, let $G_{ij}^W({\bm y};{\bm x})$ be the 
Green's tensor for the Stokes problem 
vanishing at infinity and on the plane boundary ${\cal B}$. Then 
\be
\langle v_j\rangle_1({\bm x}|\bm y)\,=\, \int_{\cal B} \left[
\langle -q \hat{n}_i+(\btau^v\cdot \hat{\bm n})_i\rangle_1(\bm s|\bm y)\,G_{ij}^W(\bm s;\bm x)+
 \langle v_j\rangle_1({\bm z}|\bm y)\,T_{ijk}^W(\bm s;\bm x)n_k\right]  d^2 s
\ee
where $T_{ijk}^W$ is the stress Green's function associated to $G_{ij}^W$ 
 and the integral is extended over the entire plane boundary \cite{KimKarrila91,Pozrikidis92} This formula can be 
considerably simplified recalling that, on the boundary, ${\bm v}$ vanishes 
everywhere outside $s$ while $G^{W}$ vanishes everywhere. 
Furthermore, on $s$, the tangential tractions also 
vanish. Hence, upon taking the $x_{1}$-axis along the normal with 
$x_1$ = 0 on the plane, we have 
\be
\langle v_j\rangle_1({\bm x}|\bm y)\,=\,
\int_s \langle v_j\rangle_1({\bm s}|\bm y)\,T_{ij1}^W(\bm s;\bm x)\, d^2 s
\label{expvj}
\ee
where now the integration is extended only over the free-slip island. 
We now consider points ${\bm x}$ such that $|\bm x-\bm s|\gg a$, but such 
that  $|\bm x-\bm s|$ is sufficiently small to be in the Stokes region  
adjacent to the boundary. It can be verified that, in this range, we have
\be
T_{ij1}^W(\bm s;\bm x)\,=\,2 T_{ij1}(\bm y;\bm x)
\left[1+O\left({a\over |\bm x-\bm s|}\right)\right]
\label{approxt3}
\ee
where $T_{ijk}$ is the free-space stress Green's function:
\be\label{freegreen}
  T_{ijk}(\bm y;\bm x)\,=\, {3\over 4\pi} {(y_{i}-x_{i})(y_{j}-x_{j})(y_{k}-x_{k}) \over |{\bm y}-{\bm x}|
^5} \, . 
\ee
Thus, (\ref{expvj}) becomes 
\be
\langle v_j\rangle_1({\bm x}|\bm y)\,\simeq\, 
2\pi a^2 T_{ij1}(\bm y;\bm x) V_i(\bm y)
\ee
where
\be
V_i(\bm y)\,=\, {1\over \pi a^2}
\int_{|\bm s-\bm y|\leq a} \langle v_i\rangle_1({\bm s}|\bm y)
\,d^2 s
\ee
is the average velocity over the disk centered at ${\bm y}$. 
Note that $V_1$ = 0 as $v_1$ = 0.
This result may now be inserted into the expression (\ref{exp39}) for the 
average field to find
\be
\langle u_j\rangle({\bm x})\,= \,u_j^0({\bm x})+2 \pi a^2\,
\int d^2y\, P({\bm y})\, T_{ij1}(\bm y;\bm x) V_i(\bm y) \, .
\label{exp4163d}
\ee
We now take the  `inner limit` of (\ref{exp4163d}) by letting the field point 
${\bm x}$ approach ${\cal B}$ to find (see e.g. 
Pozrikidis \cite{Pozrikidis92} pp. 23 and 27) 
\be
\lim_{x_{1} \rightarrow 0} T_{ij1}(\bm y;\bm x) \,=\, {1\over 2} \delta_{ij} 
\delta(\bm x -\bm y)
\label{lim3d}
\ee 
so that 
\be
\langle {\bm u}_\parallel\rangle({\bm x})\,= \, \pi a^2 
 P({\bm x})\, {\bm V}(\bm x)
\label{exp416}
\ee
where ${\bm u}_\parallel$ is the velocity component parallel to the boundary. 
Since the problem is linear, a dimensionless tensor $W_{ij}$ must exist 
such that 
\be
   V_i \,=\, {a\over \mu} W_{ij} \left(\langle \btau^w\rangle_1\cdot \hat{\bm n}
\right)_j 
\label{defw}
\ee
so that the average field satisfies the partial slip condition 
\be
\langle {\bm u}_\parallel\rangle({\bm x})\,= \,{\pi a^3\over \mu} 
 P({\bm x})\, {\sf W}\cdot \left(\langle \btau^w\rangle_1\cdot \hat{\bm n}
\right) \,\simeq \,{ a \beta\over \mu} 
\, {\sf W}\cdot \left(\langle \btau^w\rangle_1\cdot \hat{\bm n}
\right) \, .
\label{effbc}
\ee 
This equation shows that the slip length is of the order of $a\beta$. 

We can now be more specific about the assumption made at the beginning of 
section \ref{formu} as to the validity of the Stokes equations near the 
wall. The condition for this assumption is evidently that the Reynolds 
number 
\be
    Re \,=\, {2a|{\bm V}|\over \nu}
\ee
with $\nu$ the kinematic viscosity, be sufficiently small. Equation 
(\ref{defw}) shows that $|{\bm V}|$ is of the order of 
$a/\mu$ times the magnitude of the wall shear stress; a precise 
result in a particular case is derived in Appendix A.

\section{First-order problem}

While exact, the result (\ref{effbc}) expresses the effective boundary
condition on the unconditionally averaged field $\langle{\bm u}\rangle$ 
in terms of the conditionally averaged wall stress 
$\langle \btau^w\rangle_1$. In order to obtain the conditionally averaged 
velocity $\langle{\bm u}\rangle_1$ necessary to evaluate this quantity, 
one would need an effective boundary condition which would involve the wall 
stress averaged conditionally with the position of two free-slip islands 
prescribed, and so on. 
This is the well-known closure problem that arises in ensemble 
averaging. An explicit solution can only be found by truncating somehow 
the resulting hierarchy of equations. 

The lowest-order non-trivial truncation can be effected with an accuracy of 
first order in the area fraction $\beta$. It is well known that, in this 
limit, the average 'incident' $\langle{\bm w}\rangle$ may be 
approximated by the unconditional average $\langle{\bm u}\rangle$ so that 
\be
\langle {\bm u}_\parallel\rangle({\bm x})\,= \,{a \beta \over \mu} 
\, {\sf W}\cdot \left[\langle \btau\rangle(\bm x)\cdot \hat{\bm n}
\right] + o(\beta) \, .
\label{exp4161}
\ee
If the density of the islands is small, since ${\bm w}$ accounts for the 
effect of all the other islands on the one centered at ${\bm y}$, 
$\langle {\bm w}\rangle_1$ is slowly varying near ${\bm y}$ 
so that
\be
 \langle {\bm w}\rangle_1(\bm x)\,=\, 
 \langle {\bm w}\rangle_1(\bm y)+[(\bm x-\bm y)\cdot 
\div]  \langle {\bm w}\rangle_1 ({\bm y}) +\ldots
\ee
and, therefore, 
\be
 \langle\tau^w_{jk}\rangle_1 \,=\, \mu \left( \pat_j \langle  w_k 
\rangle_1 +  \pat_k  \langle w_j \rangle_1\right) \, \simeq \,
\mu \left( \pat_j \langle u_k\rangle +  \pat_k  \langle u_j 
\rangle\right) \, = \, \langle\tau_{jk}\rangle
\ee
is approximately constant over the island $|\bm x -\bm y|\leq a$. The velocity 
field $\langle {\bm v}\rangle_1$ is therefore the solution of the Stokes 
equations (\ref{stokeq}) vanishing at infinity and whose normal component 
vanishes on the entire plane; the two tangential components vanish for 
$|\bm x -\bm y|> a$ while, for $J$ = $2,3$ and $|\bm x -\bm y|< a$ 
\be
 {\bm t}_J\cdot \left(\left[\div \langle {\bm v}\rangle_1+(  
\div \langle {\bm v}\rangle_1)^T\right]\cdot \hat{\bm n} \right) \,=\, -
{1\over \mu} {\bm t}_J \cdot (\langle \btau \rangle\cdot \hat{\bm n}) \, = \mbox{const.}
\ee
This problem is solved in the Appendix A where it is shown that
\be
W_{ij}\,=\, {8\over 9 \pi} \delta_{ij}
\ee
so that the effective boundary condition (\ref{effbc}) becomes
\be
  \langle {\bm u}_\parallel\rangle({\bm x}) \,=\, {8\over 9\pi} {a\over \mu} 
\beta({\bm x})  (
\langle {\bm \tau} \rangle({\bm x})\cdot \hat{\bm n}) +o(\beta) \, . 
\ee

It may be expected that, if the islands had an intrinsic direction (e.g., 
an elliptical shape) and were not randomly oriented, the tensor $W_{ij}$ 
would not be isotropic so that the average surface traction and surface 
velocity would not be collinear.

\section{The two-dimensional case} 

The previous analysis can also be applied to the analogous two-dimensional 
case, i.e. a surface with a random distribution of parallel, or nearly 
parallel, free-shear  strips of width $a$ oriented perpendicular to the 
flow direction. The developments at the beginning of section \ref{effbcs} are 
still valid and we may start from (\ref{expvj}) noting that, in place of (\ref{freegreen}), we have 
\be
  T_{ijk}({\bm y};{\bm x})\,=\, {1\over \pi} {(y_{i}-x_{i})(y_{j}-x_{j})(y_{k}-x_{k}) \over |{\bm y}-{\bm x}|
^4} \, . 
\label{approxt2}
\ee
so that (\ref{expvj}) becomes, in this case,  
\be
\langle v_j\rangle_1({\bm x}|\xi)\,\simeq\, 
2 a T_{j21}(\xi;\bm x) V_2(\xi) \, ,
\ee
where $\xi$ is the coordinate in the direction parallel to the plane. 
Here
\be
V_2(\xi)\,=\, {1\over a}
\int_{|\zeta-\xi|\leq a} \langle v_2\rangle_1(\zeta|\xi)
\,d\zeta
\ee
is again the average velocity over the strip centered at $\xi$. 
The expression (\ref{exp39}) for the average field is modified to 
\be
\langle u_j\rangle({\bm x})\,= \,u_{0j}({\bm x})+  2 a\,
\int d\xi\, P(\xi)\, T_{j21}(\xi;\bm x) V_2(\xi) \, . 
\label{exp416}
\ee
The analog of (\ref{lim3d}) is still valid so that 
\be
\langle  u_2 \rangle(\xi)\,= \, a \,  P(\xi)\,  V_2(\xi)
\label{exp417}
\ee
where $u_2$ is the velocity component parallel to the boundary. 

As before, from the linearity of the problem we deduce the existence 
of a dimensionless quantity $W$ such that 
\be
   V \,=\, {a\over \mu} W \langle \tau_{xy}^w\rangle_1 
\label{defw2}
\ee
so that the average field satisfies the partial slip condition 
\be
\langle  u\rangle(\xi)\,= \,{a\over \mu} \beta(\xi) 
\, W  \langle \tau_{xy}^w\rangle_1 
\label{effbc2}
\ee
where we have used the fact that the fraction of the boundary covered by the 
free-slip strips is now given by 
\be
  \beta(\xi)\,=\,\int_{|\zeta-\xi|\leq a}P(\zeta) \,d\zeta \, \simeq \, 
 a P(\xi) + O(a^2/L^2) \, .
\ee

The solution of the problem in the dilute limit is given in Appendix B. 
One find 
\be
  W \,=\, \frac{\pi}{16}
\ee
so that the effective boundary condition becomes 
\be
\langle  u\rangle(\xi)\,= \frac{\pi}{16} \,{a\over \mu} \beta(\xi) \langle \tau_{xy}\rangle + o(\beta)
\label{effbc2d}
\ee

\section{Conclusions}
\label{concl}

We have derived an effective velocity boundary condition on a wall covered 
by a random arrangement of free-slip disks or two-dimensional strips. 
For the case of disks we have found that, to leading order in the fraction 
$\beta$ of the unit area covered by the disks, the velocity satisfies a 
Navier partial slip condition with a slip length $\ell$ given by
\be
   \ell \,=\, {8\over 9 \pi} \beta\, a 
\label{slipl}
\ee
where $a$ is the common radius of the disks. 

One of the motivations of this study was the possibility that gaseous 
structures attached to the solid wall, such as nano-bubbles, 
could furnish a mechanism explaining the 
partial slip observed by several investigators and it is therefore 
interesting to examine how the result (\ref{slipl}) compares with available 
data. A full comparison would require simultaneous data for $\ell$, $\beta$ 
and $a$. The only paper in which all this information is available seems to 
be the study by Watanabe et al. \cite{Watanabeetal99}, whose data, according to Lauga \& Stone \cite{LaugaStone03} imply a slip length of about 
450 $\mu$m and an area fraction $\beta\simeq$ 10\%. With these data, 
(\ref{slipl}) gives $a\simeq$ 13 $\mu$m. Rather than disks as in the present 
study, the free-slip islands in Watanabe et al.'s work were cracks with a 
width of about 10 $\mu$m and a length of the order 100 $\mu$m. If an 
equivalent radius is estimated as $\pi a^2 = 10\times 100\,\,\mu$m$^2$, one 
finds $a\simeq$ 18 $\mu$m which is not too far from the estimate obtained 
from (\ref{slipl}). 

The study of Simonsen et al. \cite{Simonsenetal04} quotes $a\simeq$ 75 nm and $\beta \simeq$ 
60\%. With these values, the estimate (\ref{slipl})  gives 
$\ell \simeq$  13 nm. Although, for such large $\beta$'s, the relation
is probably not very accurate, this value for the slip length is in the 
ballpark measured by several investigators, such as Zhu \& Granick \cite{ZhuGranick02},  who report $0\leq \ell < 40$ nm for water, and Craig et al. \cite{Craigetal01}, who report 0 $\leq \ell < 18$ nm, for water-sucrose solutions.

Wu, Zhang, Zhang, Li, Sun, Zhang, Li \& Hu \cite{Wuetal05} measure a very low 
nano-bubble number density of about 3 bubbles per 10 $\mu$m$^2$, with 
typical radii of the order of 100 nm, which gives $\beta \simeq$ 1\%  
and $\ell\simeq$ 0.3 nm. This is small, but not out of line with some 
of the existing measurements.

The radius of surface nano-bubbles reported by Holmberg et al. \cite{Holmbergetal03} is  
in the range 25 to 65 nm while that reported by Ishida et al. \cite{Ishidaetal00} is 
of the order of 300 nm. With an area coverage of 20\%, we can
estimate a slip length between about 2 and 20 nm. Again, these numerical 
values are in the expected range. 

Tyrrell \& Attard \cite{TyrrellAttard01} and Steitz et al. \cite{Steitzetal03} measure an area coverage of 
the order of 90\%, which falls well outside the domain of applicability of 
our result. Unfortunately, neither group measured the slip length. 

Tretheway \& Meinhart \cite{TrethewayMeinhart02}  measured a slip 
length of about 1 $\mu$m, but made no estimates of area coverage or bubble 
size. With $\ell$ = 1 $\mu$m, (\ref{slipl}) gives a bubble radius 
$a$ as large as 3.5 $\mu$m  even for $\beta\sim$ 1, and larger still for 
smaller $\beta$. This is another case for which it would be of great interest 
to have some information on the surface structures. 

It is also of interest to compare our results with those of Lauga \& Stone 
\cite{LaugaStone03} obtained for flows in a tube with a periodic arrangement of 
free-slip rings perpendicular to the flow. For large tube radius, this 
arrangement should be comparable to our two-dimensional analysis. Their 
solution is numerical, but they provide an approximate analytic expressions 
valid for large tube radius, namely 
\be
 \ell\,=\, {H\over 2\pi} \,\log\left({\rm sec}\left({\pi \over 2}
\beta\right)\right)
\ee
where $H$ is the spatial period. Upon expanding for small $\beta$, we find
\be
  \ell \,\simeq\, {\pi\over 16}H \beta^2 
\ee
which, with the identification $H\beta = a$, is in precise agreement with 
our two-dimensional result (\ref{effbc2d}). Lauga \& Stone also give a 
similar result for free-slip strips parallel to the flow, but this situation 
is not comparable with either one of the two that we have considered.

\begin{acknowledgments}

We are indebted with Dr. S.M. Dammer for directing us to many pertinent 
references. M.S. is grateful to Prof. D. Lohse for several 
enlightening discussions and to STW (Nanoned programme) for financial support.

\end{acknowledgments}

\appendix

\section{Solution of the three-dimensional problem}
\label{appa}

We take the center of the island as the origin, with the $z$-axis normal 
to the plane and the $x$-axis parallel to the tangential component of 
the traction $\langle \btau \rangle\cdot{\bm n}$. 
Since the 
normal velocity component vanishes, with this choice of coordinates we 
require 
\be\label{stresscon} 
 {\pat v_x\over \pat z}\, =\, S \, , \qquad 
  {\pat v_y\over \pat z}\, =\, 0 \, , 
\ee
where
\be
    S \,=-\frac{1}{\mu} \left(\langle \btau \rangle\cdot{\bm n}\right)_x. 
\ee
Here and in the following we write ${\bm v}$ in place of 
$\langle {\bm v}\rangle_1$ for convenience. Furthermore we measure lengths 
with respect to the island radius $a$, although no special notation 
will be used to indicate dimensionless variables.  It is convenient to adopt 
a system of cylindrical coordinates $(r,\theta,z)$ 
in which $v_x\,=\, v_r\cos \theta - v_\theta \sin \theta$, 
$v_y\,=\, v_r\sin \theta + v_\theta \cos \theta$, in terms of which the 
condition (\ref{stesscon}) becomes, after suitable non-dimensionalization, 
\be
  {\pat v_r \over \pat z}\,=\, S \cos \theta \, , \qquad 
  {\pat v_\theta \over \pat z}\,=\, - S \sin \theta \, .
\label{walcd}
\ee

Following Ranger \cite{Ranger78} (see also \cite{Smith87,Davis91}), we represent 
the velocity field in the form 
\be
  {\bm v}\,=\, \div \times \left[ {\sin \theta \over r}\chi(r,z) \hat{\bm e}_z +
\div \times\left( {\cos \theta \over r}\psi(r,z) \hat{\bm e}_z\right)\right]
\label{rang}
\ee
where $\hat{\bm e}_z$ is a unit vector normal to the plane and 
\be
 {\sf L}\chi \,=\,0 \, \qquad  {\sf L}^2\psi \,=\,0 
\label{eqpc}
\ee
with
\be
 {\sf L} \,=\, {\pat^2\over \pat r^2}-{1\over r}{\pat \over \pat r}+
{\pat^2\over \pat z^2}.
\ee
The Cartesian velocity components follow from (\ref{rang}) as 
\be\label{vx}
v_{x}(r,z,\theta)=\frac{1}{2}r \partial_{r}\left[\frac{1}{r^2}\left(\partial_{z}\psi-\chi \right) \right]\mbox{cos}~2\theta+\frac{1}{2r}\partial_{r} \left(\partial_{z}\psi+\chi \right) 
\ee
\be\label{vy}
v_{y}(r,z,\theta)=\frac{1}{2}r \partial_{r}\left[\frac{1}{r^2}\left(\partial_{z}\psi-\chi \right) \right]\mbox{sin}~2\theta
\ee
\be\label{vz}
v_{z}(r,z,\theta)=-\partial_{r}\left(\frac{1}{r}\partial_{r}\psi \right) \mbox{cos}~\theta
\ee
while, from the Stokes equation, the pressure is found as 
\be
 p(r,z,\theta)\,=\, \mu {\cos \theta \over r}{\pat \over \pat z}{\sf L}\psi \,.
\ee

The solution of (\ref{eqpc}) is sought in the form of Hankel transforms 
with the result
\be
\psi \,=\, rz \int_0^\infty e^{-kz} J_1(kr)\, \tilde{\psi}(k) \, dk
\label{expsi}
\ee
\be
\chi \,=\, r \int_0^\infty e^{-kz} J_1(kr)\, \tilde{\chi}(k) \, dk \, .
\label{exchi}
\ee
The functions $\tilde{\psi}$ and $\tilde{\chi}$ must be determined by 
imposing the boundary conditions. Upon substituting (\ref{expsi}) and 
(\ref{exchi}) into (\ref{vx}) and (\ref{vy}), we find that the no-slip 
condition outside the disk is satisfied  provided that 
\be 
\displaystyle\int_{0}^{\infty} J_{1}(kr)\left( \psi(k)+\chi(k) \right) dk=\frac{d}{r}  \hspace{.2in} r>1  
\ee
\be
\displaystyle\int_{0}^{\infty} J_{1}(kr)\left( \psi(k)-\chi(k) \right) dk=0 \hspace{.2in}  r>1  
\ee
where $d$ is an integration constant to be determined later. The stress 
condition (\ref{stresscon})  inside the disk is satisfied provided that 
\be
\displaystyle\int_{0}^{\infty} J_{1}(kr)\left( -2\psi(k)-\chi(k) \right)k~dk= S r  \hspace{.2in}  0<r<1  
\ee
\be
\displaystyle\int_{0}^{\infty} J_{1}(kr)\left( -2\psi(k)+\chi(k) \right)k~dk=
b r \hspace{.2in}  0<r<1    
\ee
where $b$ is another integration constant. Upon adding and subtracting, 
we find two pairs of dual integral equations for $\psi$ and $\chi$:
\be
 \int_0^\infty J_1(kr)\tilde{\psi}(k)\,dk\,=\, {d \over 2 r} \,
\qquad 1<r
\ee
\be
 \int_0^\infty J_1(kr)\tilde{\psi}(k)\,dk\,=\, -{1 \over 4 }(b+S)r \,
\qquad 0<r<1
\ee
and 
\be
 \int_0^\infty J_1(kr)\tilde{\chi}(k)\,dk\,=\, {d \over 2 r} \,
\qquad 1<r 
\ee
\be
 \int_0^\infty J_1(kr)\tilde{\chi}(k)\,dk\,=\, {1 \over 2 }(b-S)r \,
\qquad 0<r<1 \, . 
\ee

Both these problems have the standard Titchmarsh form 
\be
 \int_0^\infty J_1(kr)\tilde{c}(k) k\,dk\,=\, {B \over r} \,
\qquad 1<r
\ee
\be
 \int_0^\infty J_1(kr)\tilde{c}(k)\,dk\,=\, Ar \,
\qquad 0<r<1
\ee
the solution of which is (see e.g. Sneddon \cite{Sneddon66} p.84)
\be 
\tilde{c}\,=\, {2\over 3}\sqrt{2\over \pi}A{J_{5/2}(k)\over \sqrt{k}}+
B{\sin k\over k}.
\ee
With this result the Hankel transforms can be evaluated in their 
complementary intervals finding 
\be
\displaystyle\int_{0}^{\infty} J_{1}(kr) \tilde{c}(k) k~dk=\left(B-\frac{4 A}{3 \pi} \right)\frac{1}{r \sqrt{r^2-1}}-\frac{4 A}{2 \pi r}\left[\sqrt{r^2-1}-r^2\mbox{arcsin}\left(\frac{1}{r}\right) \right]  \hspace{.2in}  r>1  
\ee
\be
\displaystyle\int_{0}^{\infty} J_{1}(kr) \tilde{c}(k)   dk=\frac{4}{3}\frac{A}{\pi}r \sqrt{1-r^2}+B\frac{1-\sqrt{1-r^2}}{r}  \hspace{.2in}  0<r<1.              
\ee
The second expression is regular at $r$ = 0 provided that
\be
B=\frac{4 A}{3 \pi}.
\ee
Upon imposing this condition on the solutions for $\psi$ and $\chi$ we
find
\be
d \,=\, -{8\over 9 \pi}S \, , \qquad b\,=\, {1\over 3}S
\ee
so that, finally,
\be
 \tilde{\psi}\,=\,\tilde{\chi}\,=\, -{4S\over 3\sqrt{2\pi}} {J_{3/2}(k)
\over \sqrt{k}} \,=\, {4S\over 3\pi}{k\cos k-\sin k\over k^3}
\ee
The velocity field inside the disk is readily calculated from these expressions
finding
\be
v_x(r,0,\theta)\,=\, -{4S\over 3\pi}\sqrt{1-r^2}  \hspace{.2in}
v_y(r,0,\theta) \,=\, 0. 
\ee
The average velocity over the disk is found from direct integration:
\be
{1\over \pi}
\displaystyle\int_{0}^{2\pi} d \theta \int_{0}^{1} r d r ~v_{x}(r,0,\theta)=- 
 \frac{8 S}{3 \pi} \int_{0}^{1} r \sqrt{1-r^2} dr \,=\, -\frac{8}{9\pi} S
\ee
while the $y$ component vanishes. Although not necessary for the solution of the problem at hand, it may be 
of interest to also show explicitly the expressions for the velocity and 
pressure fields away from the disk. With the definitions:   
\be
\ell_{1}=\frac{1}{2}[\sqrt{(r+1)^2+z^2}-\sqrt{(r-1)^2+z^2}]
\ee
\be
\ell_{2}=\frac{1}{2}[\sqrt{(r+1)^2+z^2}+\sqrt{(r-1)^2+z^2}]
\ee
the integrals can be evaluated to find 
(see Gradshteyn \& Ryzhik \cite{GradshteynRyzhik00} sections 6.621, 6.751 and 6.752)
\begin{eqnarray}
v_{x}(r,z,\theta)=-\frac{2 S z}{3 \pi}r^2\frac{\sqrt{\ell^{2}_{2}-1}}{(\ell^{2}_{2}-\ell^{2}_{1})\ell^{4}_{2}} \mbox{cos} 2\theta-\frac{4 S}{3 \pi}\left( \sqrt{1-\ell^{2}_{1}}-z\mbox{arcsin}\left(\frac{1}{\ell_{2}} \right) \right)\nonumber \\
-\frac{2 S z}{3 \pi}\left( \frac{\sqrt{\ell^{2}_{2}-1}}{(\ell^{2}_{2}-\ell^{2}_{1})}-\mbox{arcsin}\left( \frac{1}{\ell_{2}}\right) \right)
\end{eqnarray}
\be
v_{y}(r,z,\theta)=-\frac{2 S z}{3 \pi}r^2\frac{\sqrt{\ell^{2}_{2}-1}}{(\ell^{2}_{2}-\ell^{2}_{1})\ell^{4}_{2}} \mbox{sin} 2\theta
\ee
\be
v_{z}(r,z,\theta)=\frac{4 S z}{3 \pi}\left( -\frac{\ell^{2}_{1}\sqrt{1-\ell^{2}_{1}}}{(\ell^{2}_{2}-\ell^{2}_{1})r} \right)  \mbox{cos} 2\theta
\ee
\be
p(r,z,\theta)=\frac{8 S \mu}{3 \pi}\left( -\frac{\ell^{2}_{1}\sqrt{1-\ell^{2}_{1}}}{(\ell^{2}_{2}-\ell^{2}_{1})r} \right)           \mbox{cos} 2\theta
\ee

\section{Solution of the two-dimensional problem}

In this case it is convenient to adopt as fundamental length ${1\over 2}a$ and a Cartesian system of coordinates with $x$ along the plane direction and $y$ along the normal. The Boundary conditions of the Stokes problem for $v_x$ and $v_y$ 
become
\be
v_{x}(x,0)=0 \qquad |x| > 1
\label{condux}
\ee
\be
\partial_{y} v_{x}(x,0)=S \qquad |x| < 1.
\label{bcon1}
\ee
\be
v_{y}(x,0)=0 \qquad -\infty < x < \infty \, . 
\label{conduy}
\ee
where
\be
S=-\frac{1}{\mu}\langle \tau _{xy} \rangle.
\ee
We introduce a stream function $\psi$ in terms of which 
\be
v_{x}(x,y)=\partial_{y} \psi \, , \qquad 
v_{y}(x,y)=-\partial_{x} \psi
\ee
and 
\be
\label{streamvort}
\omega=\partial_{y} v_{x}-\partial_{x} v_{y}=\Delta \psi.
\ee
The vorticity $\omega$ is harmonic and can be written as a Fourier 
integral in the form 
\be
\omega(x,y)=\displaystyle\int_{-\infty}^{\infty}dk \exp (ikx) 
\tilde{\omega}(k)e^{-|k|y}.
\ee
By introducing the Fourier transform $\tilde{\psi}(k,y)$ of the stream 
function, substituting into (\ref{streamvort}), and integrating, we find 
\be
\tilde{\psi}(k,y)=- \frac{y \tilde{\omega}(k)}{2|k|}  e^{-|k|y} 
\ee
after elimination of an integration constant on the basis of (\ref{conduy}).
With this result, the boundary condition (\ref{condux}) becomes 
\be
\displaystyle\int_{-\infty}^{\infty}dk \exp(ikx) \tilde{\omega}(k)=S 
\qquad |x|<1
\label{bcstr}
\ee
and (\ref{bcon1})
\be
\displaystyle\int_{-\infty}^{\infty}dk \exp(ikx)  \frac{\tilde{\omega(k)}}{|k|}=0 \qquad |x|>1 \, .
\label{bcvel}
\ee
Upon writing (\ref{bcstr}) for $x$ and -$x$ and adding or subtracting, 
we find 
\be
\displaystyle\int_{-\infty}^{\infty}dk \cos(kx) \tilde{\omega}(k)=S 
\qquad 0<x<1
\label{bcstrc}
\ee
\be
\displaystyle\int_{-\infty}^{\infty}dk \sin(kx)  \tilde{\omega}(k)=0 
\qquad 0<x<1.
\label{bcstrs}
\ee
Proceeding in a similar way with (\ref{bcvel}) we have 
\be
\displaystyle\int_{-\infty}^{\infty}dk \cos(kx) \frac{ \tilde{\omega}(k)}
{|k|}=0 \qquad  1<x
\label{bcvelc}
\ee
\be
\displaystyle\int_{-\infty}^{\infty}dk \sin(kx) \frac{ \tilde{\omega}(k)}
{|k|}=0 
\qquad 1<x.
\label{bcvels}
\ee
If in (\ref{bcstrc}) we separate the integration range into $-\infty < k <0$
and $0<k<\infty$ we find
\be
\displaystyle\int_0^{\infty}dk \cos(kx)  \tilde{\omega}_+=S \qquad 0<x<1 
\qquad  \tilde{\omega}_+\,=\, \tilde{\omega}(k)+\tilde{\omega}(-k)
\label{bcstrcp}
\ee
whereas (\ref{bcstrs}) gives 
\be
\displaystyle\int_{0}^{\infty}dk \sin(kx) \tilde{\omega}_-=0 \qquad 0<x<1 
\qquad \tilde{\omega}_-\,=\,\tilde{\omega}(k)-\tilde{\omega}(-k).
\label{bcstrsm}
\ee
Similarly
\be
\displaystyle\int_0^{\infty}dk \cos(kx) \frac{\tilde{\omega}_+}{k}=0 
\qquad  1<x
\label{bcvelcp}
\ee
\be
\displaystyle\int_0^{\infty}dk \sin(kx) \frac{\tilde{\omega}_-}{k}=0
\qquad 1<x.
\label{bcvelsm}
\ee
Since the problem for $\tilde{\omega}_-$ is completely homogeneous, this 
quantity must vanish so that $\tilde{\omega}(k)$ is even in $k$ and, 
therefore, real. We are thus led to the pair of dual integral equations 
\be
\displaystyle\int_0^{\infty}dk \cos(kx)  \tilde{\omega}={1\over 2} S 
\qquad 0<x<1 
\label{intin}
\ee
\be
\displaystyle\int_0^{\infty}dk \cos(kx) \frac{\tilde{\omega}}{k}=0 
\qquad  1<x.
\label{intout}
\ee
This is a standard problem with the solution (see e.g. Sneddon \cite{Sneddon66} p. 84) 
\be
  \tilde{\omega}(k)\,=\, \frac{1}{2}S J_{1}(k) 
\ee
from which the velocity on the boundary follows as 
\be
\label{UX}
v_{x}(x,0)=-\frac{S}{2}\,\cos\left({\rm arcsin}\,x\right) \,=\, -\frac{S}{2}\sqrt{1-x^2} \qquad x<1
\ee
so that 
\be
{1\over 2}
\displaystyle\int_{-1}^{1} v_{x}(x,0) dx=-\frac{ \pi}{8}S.
\ee
This result coincides with that derived by different means in Philip \cite{Philip72}. As before, it may be of some interest to show the explicit results for the  velocity and pressure fields. One has 
\be
v_x(x,y)\,=\,-{S\over 2}\displaystyle\int_{-\infty}^{\infty}dk\,\, 
\mbox{cos} (kx) 
\left( \frac{1}{|k|}- y \right)\,\tilde{\omega}(k) e^{-|k|y}
\label{resux}
\ee
\be
v_y(x,y)\,=\,-{S\over 2}\displaystyle\int_{-\infty}^{\infty}dk\,\, 
\mbox{sin} (kx) y \tilde{\omega}(k) e^{-|k|y}
\label{resuy} 
\ee
\be
p(x,y)\,=\,-2 S\displaystyle\int_{0}^{\infty}dk \,\, \mbox{sin}(k x)
\tilde{\omega}(k)\, e^{-ky}.
\ee
The integrals can be evaluated to find 
\be
v_{x}(x,y)=-\frac{S}{2}R(x,y)-\frac{yS}{2}\partial_{y}R(x,y)
\ee
\be
v_{y}(x,y)=\frac{yS}{2}\partial_{y}I(x,y)
\ee
\be
p(x,y)=S \partial_{y} R(x,y)
\ee
with
\be
R(x,y)=-y+\sqrt{\frac{(1-x^2+y^2)+\sqrt{(1-x^2+y^2)^2+4x^2y^2}}{2}} 
\ee
\be
I(x,y)=x+\sqrt{\frac{-(1-x^2+y^2)+\sqrt{(1-x^2+y^2)^4+4x^2y^2}}{2}} \, .
\ee

\end{document}